\documentclass{jps-cp}

\usepackage{txfonts}   
\usepackage{booktabs}
\usepackage{multirow}
\usepackage{float}
\usepackage{array}
\usepackage{multirow}

\title{Structural Anisotropy in Bi$_2$Se$_3$/$\lambda$-(BETS)$_2$GaCl$_4$ Topological Insulator/Superconductor Bilayer}

\author{Fathan M. \textsc{Santana}$^{1*}$,
Arifa D. \textsc{Agustin}$^{1,2}$,
Aulia A. \textsc{Firdaus}$^{1}$,
Malik A. \textsc{Baqiya}$^{3}$,
Darminto~\textsc{Darminto}$^{3}$,
Muhammad K. \textsc{Nuryadin}$^{4}$,
Takahiko \textsc{Sasaki}$^{5}$,
Isao \textsc{Watanabe}$^{2,6}$,
Takuya~\textsc{Kobayashi}$^{6}$,
Hiromi \textsc{Taniguchi}$^{7}$,
Yasuyuki~\textsc{Ishii}$^{1}$,
Pasquale \textsc{Marra}$^{8}$,
Paolo \textsc{Mele}$^{1,9}$,
and~Dita~P.~\textsc{Sari}$^{1,2,9}$}

\inst{$^{1}$Graduate School of Engineering and Science, Shibaura Institute of Technology (SIT), Saitama, 337-8570, Japan\\
$^{2}$Nuclear Structure Research Group, RIKEN Nishina Center, Wako, 351-0198, Japan\\
$^{3}$Department of Physics, Institut Teknologi Sepuluh Nopember, Surabaya, 60111, Indonesia\\
$^{4}$Comprehensive Research Organization for Science and Society, Tokai, Ibaraki, 319-1106, Japan\\
$^{5}$Institute for Materials Research, Tohoku University, Sendai, 980-8577, Japan\\
$^{6}$Graduate School of Science, Hokkaido University, Sapporo, 060-0810, Japan\\
$^{7}$Graduate School of Science and Engineering, Saitama University, Saitama, 338-8570, Japan\\
$^{8}$Department of Engineering and Applied Sciences, Sophia University, Tokyo, 102-8554, Japan\\
$^{9}$Innovative Global Program, College of Engineering, SIT, Saitama, 337-8570, Japan}

\email{na26108@shibaura-it.ac.jp}

\recdate{September 25, 2026}

\abst{We investigate the crystallographic-orientation dependence of the response of $\lambda$-(BETS)$_2$GaCl$_4$ ($\lambda$-Ga) to Bi$_2$Se$_3$ deposition. Cross-sectional focused-ion-beam (FIB) imaging confirms formation of a deposited Bi$_2$Se$_3$-containing layer and a distinct interface with $\lambda$-Ga. Single-crystal X-ray diffraction (SC-XRD) shows that the coated $\lambda$-Ga crystal remains structurally intact but exhibits a strongly orientation-dependent, non-monotonic response, with the largest modification for the $ab$-plane after 4 min of deposition, whereas the $ac$-plane remains comparatively unchanged. This condition corresponds to the intermediate-deposition regime previously identified at a Bi$_2$Se$_3$ thickness of approximately 110 nm. First-principles calculations based on the refined structures show the largest orientation-dependent electronic modification under the same conditions. Complementary optical spectroscopy of Bi$_2$Se$_3$/SiO$_2$ films shows deposition-time-dependent evolution of the film response. These results establish a pronounced crystallographic anisotropy in the structural and calculated electronic response of $\lambda$-Ga to Bi$_2$Se$_3$ deposition.}

\kword{Bi$_2$Se$_3$, $\lambda$-(BETS)$_2$GaCl$_4$, topological insulator, molecular superconductor, crystallographic orientation, optical conductivity}

\begin{document}
\maketitle

\section{Introduction}
Topological-insulator/superconductor (TI/SC) heterostructures provide a platform for exploring the interplay between superconductivity and topological surface states, including the possibility of proximity-induced topological superconductivity and Majorana-related states~\cite{Marra2024,FuKane2008,Zareapour2012,Wang2013}. Bi$_2$Se$_3$ is particularly suitable for such hybrid structures because of its well-established Dirac surface state and layered van der Waals structure~\cite{ZhangKaneMele2012,Nakajima1963}. Proximity-induced superconductivity has been demonstrated in Bi$_2$Se$_3$ on conventional inorganic superconductors~\cite{Liu2014}, whereas interfaces with molecular superconductors, whose electronic structures are highly sensitive to molecular packing, remain largely unexplored.

The molecular superconductor $\lambda$-(BETS)$_2$GaCl$_4$ ($\lambda$-Ga) is an attractive counterpart because of its pronounced electronic and superconducting anisotropy, including an anisotropic nodal superconducting gap~\cite{Sari2021} and a proposed high-field FFLO state~\cite{Imajo2021}. In $\lambda$-Ga, BETS molecules stack along the crystallographic $a$ direction, forming dimers with large intradimer transfer integrals; further molecular organization produces the characteristic packing within the unit cell~\cite{Aizawa2018}. The electronic structure is therefore sensitive to crystallographic orientation, so a Bi$_2$Se$_3$/$\lambda$-Ga heterostructure offers a degree of freedom absent from most TI/SC systems: the orientation of the coated molecular-superconductor surface.

In our previous study, the strongest structural and calculated electronic response of $\lambda$-Ga occurred after 4 min of Bi$_2$Se$_3$ deposition, corresponding to a directly measured thickness of approximately 110 nm~\cite{Santana2026}. Here, we examine whether this response depends on the orientation of the coated surface by comparing the $ab$- and $ac$-plane configurations using single-crystal X-ray diffraction (SC-XRD) and first-principles calculations, complemented by cross-sectional and optical characterization of the deposited layer.

\section{Methods}

\subsection{$\text{Bi}_2\text{Se}_3$ Thin Film Deposition and Characterization}
Bi$_2$Se$_3$ was deposited on SiO$_2$ and $\lambda$-Ga by pulsed-laser deposition using a LOTIS-TII YG-PLAD-241 system at room temperature; detailed deposition conditions and thickness characterization are reported separately~\cite{Santana2026}. Optical reflectivity of pristine $\lambda$-Ga and Bi$_2$Se$_3$/SiO$_2$ films was measured at Tohoku University over $800$--$35000~\mathrm{cm}^{-1}$ using Jasco FTIR and grating spectrometers. Optical conductivity was obtained by Kramers--Kronig (KK) transformation; identical low-frequency Drude and high-frequency $\omega^{-4}$ extrapolations were used for the 2-, 5-, and 20-min films~\cite{Dressel2004,DresselGruner2002}.

\subsection{Cross-Sectional, Structural, and Electronic Characterization}
Cross-sectional focused-ion-beam (FIB) images were obtained using a Hitachi MI-4050 system. SC-XRD of pristine and Bi$_2$Se$_3$-coated $\lambda$-Ga was performed using a Bruker D8 QUEST and analyzed with APEX5. Electronic structures were calculated using VASP 6.4~\cite{Kresse1996b} with the PAW method ~\cite{Kresse1999}, the PW91-GGA functional~\cite{Perdew1992}, a 500 eV cutoff, and a $4\times4\times4$ $k$-point mesh. The experimentally refined lattice parameters and atomic coordinates were used directly without ionic relaxation. Calculations were performed on the RIKEN HOKUSAI supercomputer.

\section{Results and Discussion}

\subsection{Optical Characterization of the Deposited Bi$_2$Se$_3$ Films}

\begin{figure}[tbh]
\centering
\includegraphics[width=1\linewidth]{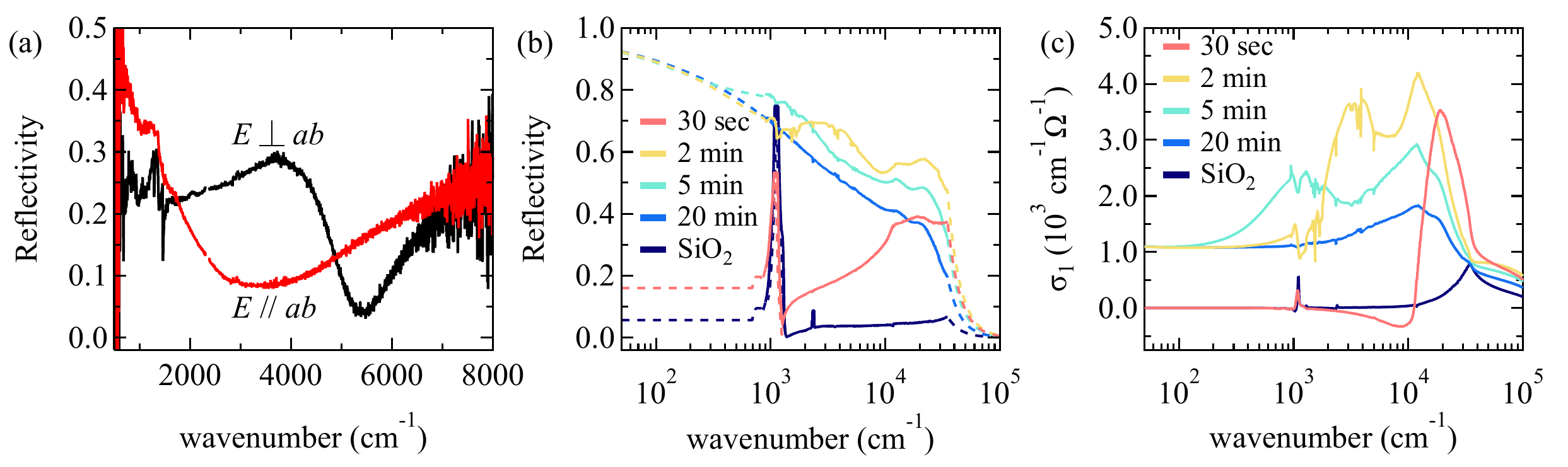}
\caption{\textbf{a:} Polarized reflectivity of pristine $\lambda$-Ga. \textbf{b:} Reflectivity spectra of films deposited on SiO$_2$ for 30~s, 2~min, 5~min, and 20~min, together with bare SiO$_2$. Dashed curves indicate the low-frequency extrapolations used for the KK analysis. \textbf{c:} Corresponding real optical conductivity $\sigma_1$ obtained from KK transformation.}
\label{Optical}
\end{figure}

Figure~\ref{Optical} shows the optical response. Pristine $\lambda$-Ga [Fig.~\ref{Optical}(a)] shows a pronounced polarization dependence, including a broad feature around $3000$--$5000~\mathrm{cm}^{-1}$ for $E\parallel ab$, close to the energy scale of the dimer charge excitation reported for $\lambda$-Ga~\cite{Aizawa2018}. In the Bi$_2$Se$_3$/SiO$_2$ films, the reflectivity changes markedly with deposition time, particularly in the low- and mid-infrared regions, and the KK-derived optical conductivity [Fig.~\ref{Optical}(c)] evolves accordingly. The apparent P1 maximum occurs near $1163~\mathrm{cm}^{-1}$ for the 5-min film and near $3526~\mathrm{cm}^{-1}$ for the 2-min film; these positions are read directly from $\sigma_1$ and are not interpreted as fitted interband transition energies. The 30-s spectrum develops nonphysical negative $\sigma_1$ values and is not used for peak assignment. The optical data therefore demonstrate a systematic evolution of the film response with deposition time, while the microscopic origin of the observed features remains unresolved.

\subsection{Cross-Sectional Characterization of the Deposited Layer}

\begin{figure}[tbh]
\centering
\includegraphics[width=0.7\linewidth]{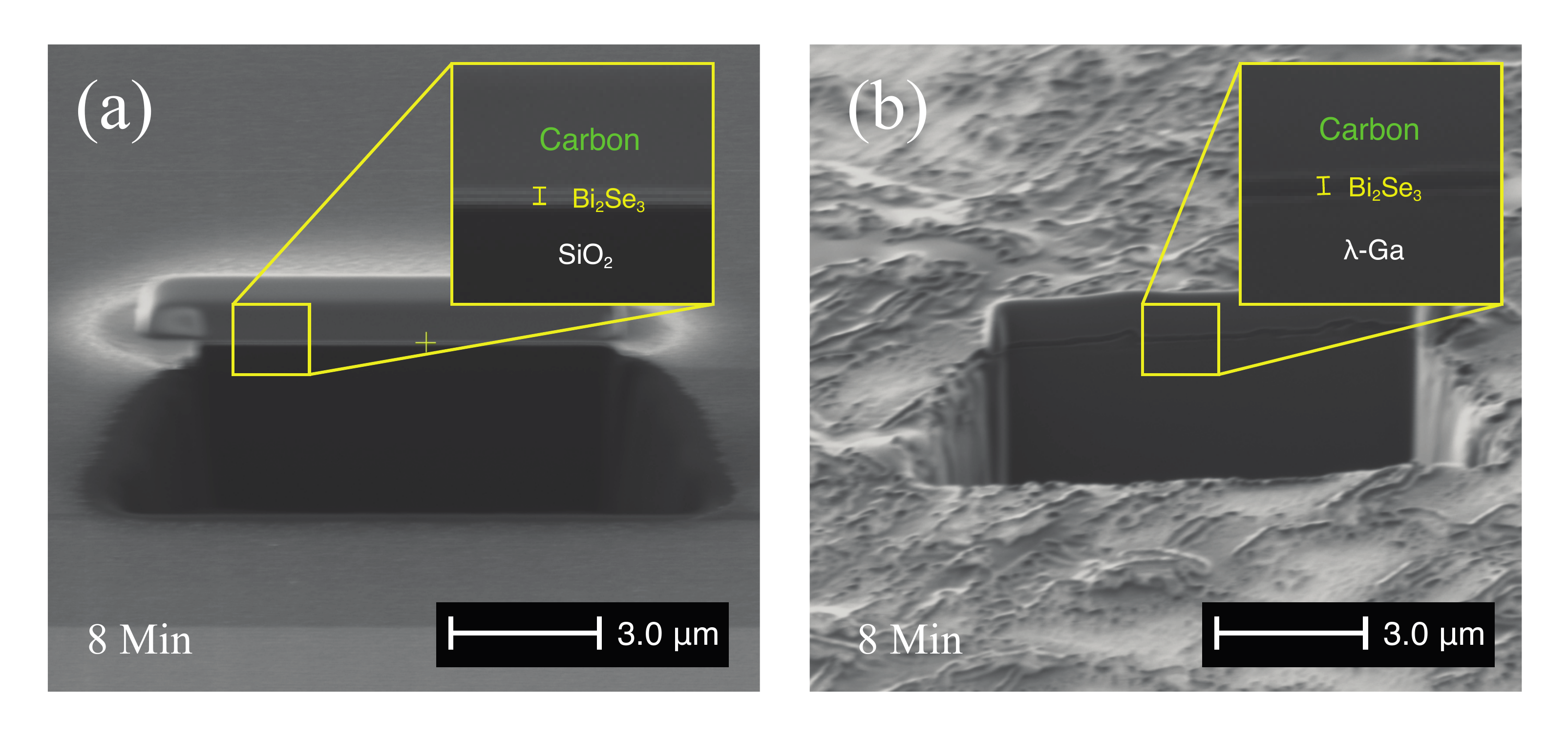}
\caption{Cross-sectional FIB images of $\text{Bi}_2\text{Se}_3$ deposited for $8\,\text{min}$ on \textbf{a:} $\text{SiO}_2$ and \textbf{b:} $\lambda\text{-(BETS)}_2\text{GaCl}_4$. The enlarged regions show the protective carbon layer, the deposited $\text{Bi}_2\text{Se}_3$-containing layer, and the corresponding substrate.}
\label{FIB}
\end{figure}

Cross-sectional FIB images of the 8-min Bi$_2$Se$_3$/SiO$_2$ and Bi$_2$Se$_3$/$\lambda$-Ga samples are shown in Fig.~\ref{FIB}. A distinct deposited layer is directly observed on both substrates. The $\lambda$-Ga cross section provides microscopic evidence of the deposited layer and the physical Bi$_2$Se$_3$/$\lambda$-Ga interface on the same type of coated crystal subsequently examined by SC-XRD. Detailed thickness characterization is reported separately ~\cite{Santana2026}.
Electron-probe microanalysis and SEM-EDS of the same deposition series show progressive incorporation of Bi and Se with deposition time~\cite{Santana2026}; the stoichiometry of the deposited layer is not determined in the present work.

\subsection{Crystallographic Integrity and Anisotropic Response to Bi$_2$Se$_3$ Deposition}

\begin{figure}[tbh]
\centering
\includegraphics[width=1\linewidth]{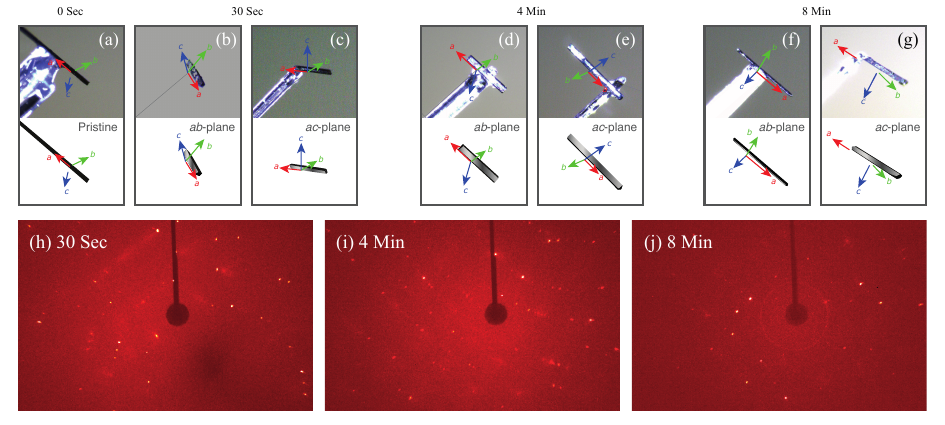}
\caption{SC-XRD setup, crystal orientations, and reflection patterns of $\lambda$-Ga crystals coated with $\mathrm{Bi}_2\mathrm{Se}_3$ by pulsed-laser deposition (PLD). \textbf{a:} Pristine $\lambda$-Ga crystal (0~s). \textbf{b,c:} 30-s, \textbf{d,e:} 4-min, and \textbf{f,g:} 8-min deposition samples oriented along the $ab$- and $ac$-planes, respectively. Red, green, and blue arrows denote the crystallographic $a$-, $b$-, and $c$-axes, respectively, with the schematics showing the corresponding crystal geometries. \textbf{h--j:} SC-XRD reflection patterns measured from the $ac$-plane after 30~s, 4~min, and 8~min of deposition, respectively. The sharp diffraction spots indicate that the single-crystalline nature is maintained across the investigated deposition times.}
\label{SC_XRD}
\end{figure}

\begin{table}[tbh]
\caption{Crystallographic data and refinement parameters of the investigated structures, all belonging to the same space group, $P\bar{1}$.}
\label{t_crystallographic}
\centering

\fontsize{6.5pt}{8pt}\selectfont
\setlength{\tabcolsep}{3.3pt} 
\renewcommand{\arraystretch}{2}

\begin{tabular}{ccccccccccccc}
\hline\hline
\textbf{Sample} & \textbf{Time} & \textbf{\begin{tabular}[c]{@{}c@{}}Coated\\Plane\end{tabular}} & \textbf{T (K)} & \textbf{$a$ (\AA)} & \textbf{$b$ (\AA)} & \textbf{$c$ (\AA)} & \textbf{$\alpha$ ($^\circ$)} & \textbf{$\beta$ ($^\circ$)} & \textbf{$\gamma$ ($^\circ$)} & \textbf{V (\AA$^3$)} & \textbf{GoF} & \textbf{\begin{tabular}[c]{@{}c@{}}Number of\\Reflections\end{tabular}} \\ \hline

$\lambda$-Ga & 0 Sec & - & 304(2) & 6.6096(2) & 16.1820(4) & 18.6205(5) & 112.5850(10) & 98.3570(10) & 96.7890(10) & 1786.01(9) & 1.097 & 6237 \\ \hline

\multirow{6}{*}{Bi$_2$Se$_3$/$\lambda$-Ga}
& \multirow{2}{*}{30 Sec} & ab & 302(2) & 6.6119(2) & 16.1819(4) & 18.6230(5) & 112.5770(10) & 98.3920(10) & 96.7480(10) & 1786.99(9) & 1.200 & 6283 \\
& & ac & 302(2) & 6.6105(2) & 16.1798(5) & 18.6262(5) & 112.5760(10) & 98.3460(10) & 96.7800(10) & 1786.82(9) & 1.184 & 6277 \\ \cline{2-13}

& \multirow{2}{*}{4 Min} & ab & 303(2) & 6.6434(16) & 16.223(5) & 18.640(5) & 112.567(11) & 98.477(12) & 96.518(12) & 1802.8(8) & 1.028 & 6346 \\
& & ac & 302(2) & 6.6112(2) & 16.1861(4) & 18.6256(4) & 112.5730(10) & 98.3810(10) & 96.7780(10) & 1787.45(8) & 1.143 & 6348 \\ \cline{2-13}

& \multirow{2}{*}{8 Min} & ab & 303(2) & 6.6197(8) & 16.200(3) & 18.647(2) & 112.591(5) & 98.350(6) & 96.718(8) & 1793.7(4) & 1.037 & 6262 \\
& & ac & 304(2) & 6.6103(3) & 16.1890(9) & 18.6258(10) & 112.609(2) & 98.366(2) & 96.777(2) & 1787.15(16) & 1.079 & 6281 \\ \hline\hline

\end{tabular}
\end{table}

Fig.~\ref{SC_XRD}(a--g) shows the experimental configurations of the $\text{Bi}_2\text{Se}_3/\lambda$-Ga samples prepared with different coating planes and deposition times, while Fig.~\ref{SC_XRD}(h--j) shows representative diffraction patterns. Sharp diffraction spots remain visible throughout the deposition series, showing that the single-crystalline $\lambda$-Ga framework is retained after Bi$_2$Se$_3$ deposition (Fig. 3). The refined structures nevertheless exhibit a clear orientation-dependent response. The $ac$-plane remains near the pristine volume of $1786.01(9)~\text{\AA}^3$, whereas the $ab$-plane changes from $1786.99(9)$ to $1802.8(8)$ and $1793.7(4)~\text{\AA}^3$ at 30 s, 4 min, and 8 min, respectively. Thus, the largest structural anisotropy occurs at 4 min. The reflection count also peaks near this condition and decreases again at 8 min, although it is treated only as a supporting descriptor.

Simulated powder X-ray diffraction (pXRD) patterns were calculated from the refined $\lambda$-Ga structures in the $2\theta$ regions $29.2$--$29.7^\circ$ and $47.2$--$48.4^\circ$, which contain the strongest Bi$_2$Se$_3$ reflection (0\,1\,5) and the (0\,0\,15)/(1\,1\,6) reflections of the standard structure~\cite{Nakajima1963} (hexagonal setting of $R\bar{3}m$), and additionally in the regions $18.3$--$18.9^\circ$ and $40.1$--$40.7^\circ$ near the (0\,0\,6) and (1\,0\,10) reflections (Fig.~4). All patterns were referenced to the $\lambda$-Ga (0\,0\,1) reflection at $2\theta \approx 5.25^\circ$, whose position varies by less than $0.01^\circ$ among the refined structures; all peaks shown are $\lambda$-Ga reflections and are not assigned to the Bi$_2$Se$_3$ overlayer. The patterns show the same trend: small orientation-dependent differences at 30 s, the largest shifts at 4 min, and partial convergence at 8 min. Together with the direct FIB observation of the interface, these results show that Bi$_2$Se$_3$ deposition is accompanied by a non-monotonic, crystallographically anisotropic modification of $\lambda$-Ga. The SC-XRD data are not depth-resolved and therefore do not determine the spatial extent of the modification from the interface.

\begin{figure}[tbh]
\centering
\includegraphics[width=1\linewidth]{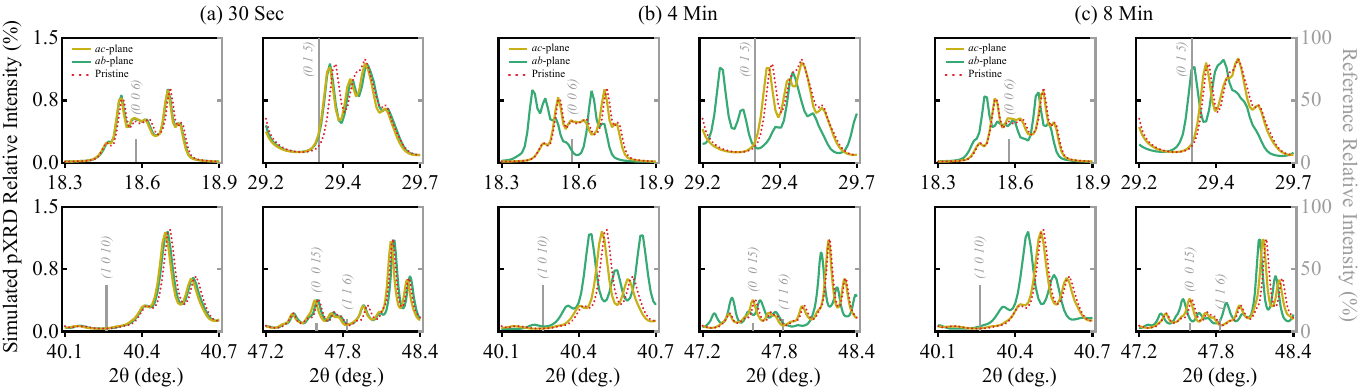}
\caption{Simulated powder diffraction patterns calculated from the refined $\lambda$-Ga structures for the \textit{ab}- and \textit{ac}-plane configurations and pristine $\lambda$-Ga after \textbf{a:} 30~s, \textbf{b:} 4~min, and \textbf{c:} 8~min of deposition, in $2\theta$ regions near the Bi$_2$Se$_3$ (0\,0\,6), (0\,1\,5), (1\,0\,10), and (0\,0\,15)/(1\,1\,6) reflections. Grey bars (right axis) mark the reflections of Ref.~\cite{Nakajima1963} with their relative intensities. All patterns are referenced to the $\lambda$-Ga (0\,0\,1) reflection; all peaks shown are $\lambda$-Ga reflections.}
\label{XRD_Pattern}
\end{figure}

\subsection{Electronic Response to the Deposition-Induced Structural Modification}

\begin{figure}[tbh]
\centering
\includegraphics[width=1\linewidth]{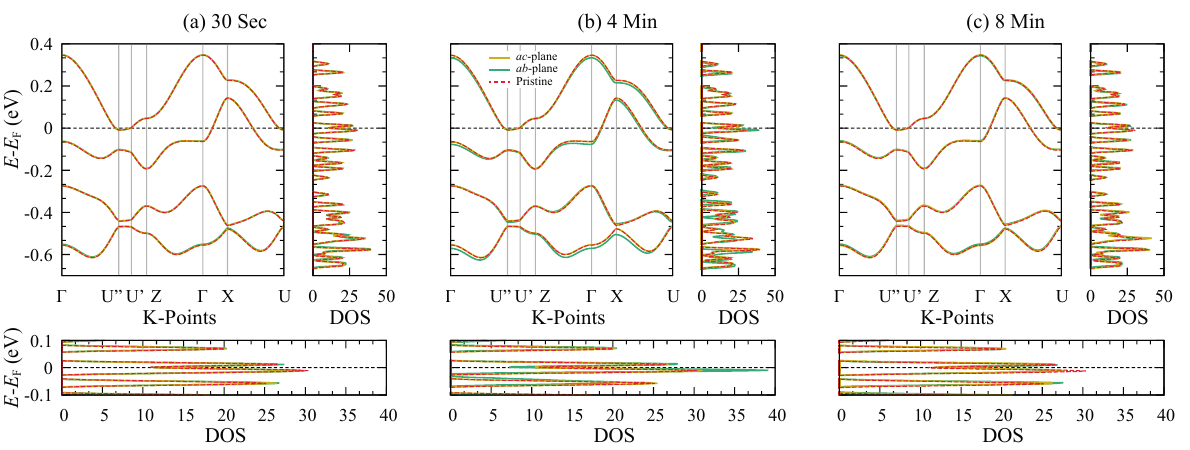}
\caption{\textbf{a:} Calculated electronic band structures and DOS of $\lambda$-Ga for the $ab$- and $ac$-plane configurations after 30~s, \textbf{b:} 4~min, and \textbf{c:} 8~min of $\mathrm{Bi}_2\mathrm{Se}_3$ deposition. Green, yellow, and red dotted curves denote the \textit{ab}-, \textit{ac}-, and pristine configurations, respectively. The largest orientation-dependent difference occurs at 4~min.}
\label{DFT}
\end{figure}

Electronic structures calculated directly from the experimentally refined $\lambda$-Ga structures are shown in Fig.~\ref{DFT}. Because the refined lattice parameters and atomic coordinates were retained without ionic relaxation, the calculated differences directly reflect the measured structural changes. Near the Fermi level, four bands originate primarily from the four BETS molecules in the unit cell. In the dimer-limit model, one electron is associated with each BETS dimer, resulting in two bands crossing the Fermi level. The $ab$- and $ac$-plane configurations are nearly identical at 30 s, differ most strongly at 4 min in both band structure and density of states (DOS) near $E_{\mathrm{F}}$, and become more similar again at 8 min. The characteristic $\lambda$-Ga frontier-band structure remains intact throughout the series.

The coincidence of the maximum structural and calculated electronic anisotropy at 4~min links the electronic modification to the measured response of the coated crystal. Together with the FIB evidence for a distinct Bi$_2$Se$_3$/$\lambda$-Ga interface, this indicates that the structural and electronic modification is induced by Bi$_2$Se$_3$ deposition. Because the calculations use the refined $\lambda$-Ga structures rather than an atomistic interface model, the microscopic mechanism, including whether the response originates at the interface itself, requires further study.

\section{Conclusion}

We have investigated the crystallographic-orientation dependence of the structural and electronic response of $\lambda$-(BETS)$_2$GaCl$_4$ to Bi$_2$Se$_3$ deposition. Cross-sectional FIB imaging directly confirms formation of the Bi$_2$Se$_3$/$\lambda$-Ga interface, while SC-XRD of the coated crystals reveals a pronounced non-monotonic and orientation-dependent structural response. The strongest modification occurs for the $ab$-plane after 4 min of deposition, corresponding to the approximately 110-nm intermediate-thickness regime identified previously, whereas the $ac$-plane remains comparatively unchanged. Simulated diffraction patterns and first-principles calculations based on the refined structures show their largest orientation-dependent differences at the same 4-min condition and partially recover at 8 min. Thus, the response of $\lambda$-Ga is not determined simply by increasing Bi$_2$Se$_3$ thickness but is maximized in a characteristic intermediate regime. Complementary optical measurements confirm deposition-dependent evolution of the Bi$_2$Se$_3$ film response, while the microscopic origin and depth profile of the interfacial modification remain subjects for future study. Determination of the stoichiometry of freshly deposited films and magnetization measurements of the 30-s and 4-min \textit{ab}-plane samples are in progress.

\section*{Acknowledgements}
FMS is supported by the SIT scholarship for foreign graduate students. DPS and IW are supported by the JSPS KAKENHI No. 21K13885, 24K06133, 25H01559, 23H04529, the JSPS Asia-Africa Core-to-Core No. JPJSCCB20240005, and JST NEXUS Y-tec No. Y2024L0906032, Y2025L1015059. Computations were performed on the RIKEN HOKUSAI (ID: RB230087). P. Ma is partially supported by the JSPS KAKENHI No. 23K13028, 20K14375, 22H05111, and 24H00826.


\end{document}